\documentclass[preprint,aps]{revtex4}

\begin{document}


\title{Absence of U(1) spin liquids in two dimensions}

\author{Igor F. Herbut and Babak H. Seradjeh}

\affiliation{Department of Physics, Simon Fraser University,
Burnaby, British Columbia, Canada V5A 1S6}

\author{Subir Sachdev}
\affiliation{Department of Physics, Yale University, P.O. Box
208120, New Haven CT 06520-8120}

\author{Ganpathy Murthy}
\affiliation{Department of Physics and Astronomy, University of
Kentucky, Lexington KY 40506-0055}

\date{June 18, 2003}

\begin{abstract}
Many popular models of fractionalized spin liquids contain neutral
fermionic spinon excitations on a Fermi surface, carrying unit
charges under a compact U(1) gauge force. We argue that instanton
effects generically render such states unstable to confinement in
two spatial dimensions, so that all elementary excitations are
gauge neutral, and there is no spinon Fermi surface. Similar
results are expected to apply to SU(2) spin liquids. However,
fractionalized states can appear when the gauge symmetry is broken
down to a discrete subgroup by the Higgs mechanism. Our argument
generalizes earlier results on confinement in the pure gauge
theory, and on the instability of the U(1) `staggered flux' and
`algebraic' spin liquids with a Dirac spectrum for fermionic
spinons.
\end{abstract}

\maketitle

\section{Introduction}
\label{sec:intro}

The anomalous normal state of the cuprate superconductors has 
focused a great deal of attention on exotic states with electron
fractionalization. One model, which has enjoyed a great deal of
popularity
\cite{pwa,ba,aff,nagaosa1,ioffe1,ki,ioffe2,ioffe3,nagaosa2,nagaosa3,russians,altshuler1,altshuler2,altshuler3,su21,su22,kimlee,wennew,ichi,su,salk,jacklee,ng},
has the electron spin reside on neutral $S=1/2$ fermionic
`spinons' which form a Fermi surface. The microscopic theory of
the formation of spinons shows that each spinon carries unit
charge of a compact gauge group, which is usually
\cite{ba,nagaosa1,ioffe1,ki,ioffe2,ioffe3,nagaosa2,nagaosa3,russians,altshuler1,altshuler2,altshuler3,kimlee,wennew,ichi,su,salk,jacklee,ng}
U(1); however, somewhat more involved theories with a SU(2) gauge
group have also been considered \cite{aff,wennew,su21,su22,salk}.
The spinons interact with each other, and also with other possible
`holon' excitations, via exchange of the gauge boson. We will
focus here on the U(1) case, but our arguments appear to have a
direct generalization to the SU(2) case.

Polyakov \cite{polyakov} argued many years ago that the pure
compact U(1) gauge theory is always confining at zero temperature
in two spatial dimensions. In other words, there is no `Coulomb'
phase with a massless `photon' excitation, and oppositely charged
static test particles experience a confining potential which
increases linearly with the distance between them. The confinement
was induced by the proliferation of `instanton' tunneling events.
Given this fundamental result, one might wonder if similar effects
are important in the models of electron fractionalization
\cite{ba}. Indeed, in theories of quantum antiferromagnets with
collinear spin correlations, such effects were crucial in
disrupting possible spin liquid states and leading to ground
states with bond order and confined spinons \cite{rs}.

Many models of spin liquids have gapless fermionic spinon
excitations, and in these cases the argument for confinement by
proliferation of instantons is more delicate. The `staggered flux'
and related `algebraic spin liquids' possess spinons with a
relativistic Dirac-like spectrum, and consequently the instantons
interact with an action which depends logarithmically on their
spacetime separation \cite{ioffelarkin}. By examining the
renormalization of the fugacity of a single instanton, several
authors \cite{ioffelarkin,rantner,wennew} have argued that this
logarithmic coupling leads to a regime in which instantons are
suppressed, and so a fractionalized spin liquid regime is stable.
Similar arguments have also been advanced recently for cases where
there are relativistic massless Bose fields \cite{sudbo}. However,
these arguments neglect the screening of the instanton-instanton
coupling by bound instanton pairs at shorter scales. A number of
other authors \cite{quote,marston,murthy,sp} have mentioned this
screening, and argued that it disrupts the binding of instantons
in 2+1 spacetime dimensions; the instantons are always ultimately
in a plasma phase, interacting via an exponentially screened
interaction, and the spinons carrying dual `electric' charges are
consequently permanently confined (these arguments are presented
here in Appendix~\ref{cg}). Recently, the case for permanent
confinement was put on a firmer footing by two of us \cite{igor1}:
an explicit duality mapping of a compact U(1) gauge theory minimally coupled
to relativistic fermions showed that the plasma state of
instantons was the only stable phase, and so there were no
`electrically' charged fermionic spinons. This result implies that the staggered flux
and algebraic spin liquids do not exist in two spatial dimensions. 

The problem addressed in this paper is the case of a compact U(1)
gauge theory coupled to {\em non-relativistic} matter, with the
fermionic spinons forming a Fermi surface. As noted above, this
case has been the focus of much attention
\cite{ba,nagaosa1,ioffe1,ki,ioffe2,ioffe3,nagaosa2,nagaosa3,russians,altshuler1,altshuler2,altshuler3,kimlee,wennew,ichi,su,salk,jacklee,ng}
in models of the normal state of the cuprates. For this case,
Nagaosa \cite{nagaosa3} has advanced arguments that the
dissipation caused by the spinon Fermi surface can suppress the
instanton quantum fluctuations, and so stabilize the spin liquid
state. However, using the method of Ref.~\onlinecite{igor1}, we
show below that Nagaosa's argument also neglects the screening
between instantons \cite{nagaosanote}, and that the spinon Fermi
surface is generically unstable in two spatial dimensions at zero
temperature.

A surviving route to fractionalization in two spatial dimensions
arises in physical situations with a Higgs field which does not
belong to the fundamental representation of the gauge group
\cite{fs}. In the spin liquid case, such a Higgs field is provided
by the pairing of the spinons \cite{z2rs,z2wen,z2mud,sf} in a BCS-like state,
and the formation of such a BCS condensate breaks the U(1) gauge
symmetry down to $Z_2$. The resulting $Z_2$ gauge theory does
permit a deconfined phase, with spinons which carry only $Z_2$
gauge charges. Note, however, that because of the pairing, there
is no spinon Fermi surface in this case either. 

For completeness, we also 
mention the `chiral spin liquid' \cite{z2wen,kalmeyer,chiral,fradkin}, which allows fractionalization
in the presence of time-reversal symmetry breaking, but again without
a spinon Fermi surface. 

This paper will only be concerned with matter in the fundamental
representation of the U(1) gauge group. Our analysis below shows that the
only possible exceptions to conventional confinement in such
situations arise at special critical points where a certain
relevant coupling in the action accidentally vanishes: at least
one parameter has to be tuned to a particular value for this to
happen. At such critical points, the gauge theory is not in a
conventional deconfined Coulomb phase either, and it is likely
that there are no quasiparticle excitations. These points are
analogous to the critical Rokhsar-Kivelson points in quantum dimer
models \cite{rk,sondhi}.

Our analysis of the compact U(1) gauge theory coupled to fermionic
spinons on a Fermi surface begins in Section~\ref{sec:duality}
with a duality mapping to a sine-Gordon-like theory. The
renormalization group analysis of the latter theory is presented
in Section~\ref{sec:rg} and Appendix~\ref{app}. Appendix~\ref{cg}
contains a discussion of similar physics expressed directly in
terms of the statistical mechanics of instantons.

\section{Duality mapping}
\label{sec:duality}

A fundamental characteristic of the proposed spin liquid state
\cite{nagaosa1,ioffe1} is the overdamped nature of the transverse
gauge field propagator. This damping arises from the low energy
fluctuations of the spinon Fermi surface. We denote the components
of the compact U(1) gauge field by $a_{\mu}$ with $\mu = 0,x,y$,
the zero'th component representing the imaginary time direction;
also by $a_i$ we denote only the spatial components of $a_{\mu}$.
Then, after integrating out the fermionic spinon fields, the
action, $\mathcal{S}$, for $a_{\mu}$ at small wavevectors ($k$)
and small frequencies ($\omega$) has the following singular form
\cite{nagaosa1,ioffe1,ki,nagaosa2}
\begin{equation}
\mathcal{S} = \int \frac{d^2 k d \omega}{8 \pi^3} \left[
\frac{1}{2} a_i a_j \left( \delta_{ij} - \frac{k_i k_j}{k^2}
\right) \left( |\omega| \sigma (k) + \chi_d k^2 \right) +
\frac{\chi}{2}\left( 1 + \frac{\gamma |\omega|}{k} \right)\left(
a_0 - \frac{\omega}{k^2} k_i a_i \right)^2 \right]. \label{e1}
\end{equation}
Here $\sigma (k)$ is the spinon ``conductivity'', $\chi_d$ is the
spinon diamagnetic susceptibility, $\chi$ is the spinon
compressibility, and $\gamma$ is a damping coefficient
characteristic of the spinon Fermi surface. All of the terms in
$\mathcal{S}$ are characteristic of a fermionic spinon system with
a finite density of states at the Fermi level (for the
relativistic case with a Dirac spinon spectrum, there is a
vanishing density of states at the Fermi level, and this makes the
effective gauge field action quite different \cite{igor1}). For
$\sigma (k)$, it is conventional to assume
\cite{nagaosa1,nagaosa2}
\begin{equation}
\sigma (k) \sim \left\{ \begin{array}{cc} \ell & \mbox{ for $k
\ell < 1$} \\
1/k & \mbox{ for $k \ell > 1$}
\end{array} \right.
\label{e2}
\end{equation}
where $\ell$ is the mean-free path associated with scattering off
static impurities.

Our subsequent analysis is aided by writing $\mathcal{S}$ in a
form which is explicitly gauge invariant. We define the
electromagnetic field $F_{\mu \nu} = p_{\mu} a_{\nu} - p_{\nu}
a_{\mu}$ where $p_0 = \omega$, $p_i = k_i$. Then, it is easy to
see that
\begin{equation}
\mathcal{S} = \int \frac{d^2 k d \omega}{8 \pi^3} \left[
\frac{1}{4} \left( \frac{|\omega| \sigma (k)}{k^2} + \chi_d +
\frac{1}{e^2} \right) F_{ij}^2
 + \frac{1}{2}\left\{ \frac{\delta_{ij}}{e^2} + \frac{\chi k_i k_j}{k^4} \left( 1 + \frac{\gamma
|\omega|}{k} \right)\right\}F_{i0} F_{j0} \right]. \label{e3}
\end{equation}
The expression (\ref{e3}) for $\mathcal{S}$ also includes a
regular $F_{\mu\nu}^2 /(4 e^2)$ term in the action for the gauge
field; such a term will invariably be generated by integrating out
high energy fluctuations of matter fields at short scales.

We may now proceed to dualize the above action for the gauge
fields. Before doing so, it is necessary to explicitly account for
the compact nature of the U(1) gauge group on the scale of
underlying lattice. We do this by discretizing space and time, and
writing the compactified lattice version of the action (\ref{e3})
as
\begin{eqnarray}
\mathcal{S} = \sum_{x,y,\tau,\tau'}\Biggl[ \frac{1}{4}
(F_{ij}(x,\tau)- 2\pi n_{ij}(x,\tau)) V_t (x-x',\tau-\tau')
(F_{ij}(x' ,\tau')-2 \pi n_{ij}(x',\tau'))  \nonumber \\
\\ + \frac{1}{2} (\Delta_i (F_{i0}(x,\tau)-2 \pi n_{i0}(x,\tau)) )
V_l(x-x',\tau-\tau') (\Delta_j (F_{j0}(x',\tau')-2 \pi
n_{j0}(x',\tau' )) )\Biggr] \nonumber \\
+ \frac{1}{2 e^2}\sum_{x,\tau}
 (F_{i0}(x,\tau)-2 \pi n_{io}(x,\tau)) ^2 \nonumber, \label{e4}
\end{eqnarray}
where the transverse and the longitudinal parts of the interaction
in the Fourier space read
\begin{equation}
V_t(k,\omega) = \frac{|\omega| \sigma(k)}{k^2} + \chi_d
+\frac{1}{e^2}, \label{e5}
\end{equation}
\begin{equation}
V_l(k,\omega) =  \frac{\chi}{k^4} \left(1+ \gamma
\frac{|\omega|}{k}\right). \label{e6}
\end{equation}
The integers $n_{\mu \nu}(x,\tau)$ serve to account for the
compact nature of the gauge fields, in the spirit of the Villain
approximation. On a lattice, the electromagnetic tensor $F_{\mu
\nu}= \epsilon_{\mu\nu\alpha}\epsilon_{\alpha\rho\sigma}
\Delta_{\rho} a_{\sigma}$, where $\Delta_\mu$ is the lattice
(discrete) derivative.

  Performing the Hubbard-Stratonovich transformation using fields $c_{\mu\nu}$ residing
on the plaquettes of the lattice, we find
\begin{eqnarray}
\mathcal{S}= \sum_{x,x',\tau,\tau'} \Biggl[ c_{ij}(x,\tau) V_t
^{-1} (x-x', \tau-\tau') c_{ij}(x',\tau') + \frac{1}{2} e^2
c_{i0}(x,\tau) \Biggl\{ \delta_{ij} \delta_{x x'} \delta_{\tau
\tau'} \nonumber \\ + \left(\frac{1}{|\Delta|^2} -
\left(\frac{1}{e^2} \delta_{x x'} \delta_{\tau \tau'} + V_l
(x-x',\tau-\tau') |\Delta|^2
 \right)^{-1} \right) \Delta_i \Delta_j \Biggr\} c_{j0} (x',\tau') \Biggr] \nonumber \\
+i \sum_{x,\tau} c_{\mu\nu}(x,\tau) (F_{\mu \nu} (x,\tau) - 2\pi n_{\mu
\nu}(x,\tau) ). \label{e7}
\end{eqnarray}
The integral over the gauge fields can now be performed exactly,
and it enforces the constraint
\begin{equation}
c_{\mu \nu} (x,\tau) = \frac{1}{2} \epsilon_{\mu \nu \alpha}
\Delta_{\alpha} \Phi(x,\tau). \label{e8}
\end{equation}
The sum over the integers $n_{\mu \nu}(x,\tau)$ can also be
performed, and it forces the new variable $\Phi(x,\tau)$ to be an
integer as well. From the constraint in Eq. (\ref{e8}) it
immediately follows,
\begin{equation}
\Delta_i c_{i0}(x,\tau) =0, \label{e9}
\end{equation}
and the corresponding part in Eq. (\ref{e7}), which includes the
longitudinal part of the interaction $V_l$, completely drops out.
Relaxing the integer constraint on $\Phi$ by introducing a small
fugacity $y$ in the usual way \cite{jose},
upon return to the continuum limit
we find the dual sine-Gordon action for the instantons
\begin{eqnarray}
\mathcal{S}_{sG} &= & \frac{1}{2} \int \frac{d^2 k d \omega}{8
\pi^3} \left[ |\Phi(k,\omega)|^2 \left( \frac{\omega^2}{|\omega|
\sigma (k)/k^2 + \chi_d + 1/e^2} + e^2 k^2 \right) \right]
\nonumber \\
&~&~~~~~~~~~~~~~~~~~ - 2 y \int d^2 x d \tau \cos(2 \pi
\Phi(x,\tau) ). \label{e10}
\end{eqnarray}
A corresponding sine-Gordon theory for the relativistic matter
case appears in Ref.~\onlinecite{sudbo}.
The problem of confinement in the original theory now reduces to
understanding the possible phases of $\mathcal{S}_{sG}$, to which
we turn in the next section \cite{comment}.

\section{Renormalization group analysis}
\label{sec:rg}

 To be specific, we assume that in the sine-Gordon theory
the wavevectors $|\vec{k}|< \Lambda$, with $\Lambda \ell <1$, and
frequencies $|\omega| < \Omega$. To simplify the calculation, we
also take $\Omega \gg (\chi_d +e^{-2})\Lambda ^2 /\ell$, so one
can write the sine-Gordon
 field propagator for low momenta and frequencies as
\begin{equation}
G^{-1} (\vec{k},\omega)= \frac{|\omega| k^2}{\ell} + a_k  k^2 + a_\omega
\omega^2, \label{e11}
\end{equation}
where the bare (unrenormalized) coefficients are $a_k  = e^2$, and
$a_\omega =0$. Our methods below are easily generalized to the
clean limit case where $\sigma (k) \sim 1/k$, and the results are
very similar.

Let us integrate out the short-distance
modes with $\Lambda/b < |\vec{k}|<\Lambda $,
and $|\omega|<\Omega$, with $\ln(b) \ll 1$. The parameters for the
remaining long-distance modes are then changed as
\begin{equation}
\ell(b) = \ell, \label{e12}
\end{equation}

\begin{equation}
a_k (b) =  a_k  + \frac{1}{2} y^2 e^{-G_> (\vec{x}=0,\tau=0)}
\int d^2 \vec{x} d\tau |\vec{x}|^2 ( e^{G_>(\vec{x},\tau)} -1) ,
\label{e13}
\end{equation}

\begin{equation}
a_\omega (b) = b^2 [a_\omega  + y^2 e^{-G_> (\vec{x}=0,\tau=0)}
\int d^2 \vec{x} d\tau \tau ^2 ( e^{G_>(\vec{x},\tau)} -1)] ,
\label{e14}
\end{equation}

\begin{equation}
y(b) = b^2 y e^{-\frac{1}{2} G_> (\vec{x}=0,\tau=0)} ,
\label{e15}
\end{equation}

where the correlation function is defined as
\begin{equation}
G_> (\vec{x},\tau) = \int_{\Lambda/b < |\vec{k}|<\Lambda }
\frac{d^2 \vec{k}}{(2\pi)^2} \int_{-\Omega} ^{\Omega} \frac{d\omega}{2\pi}
e^{i\vec{k}\cdot \vec{x} + i\omega \tau} G(\vec{k},\omega).
\label{e16}
\end{equation}

   A crucial feature of the above recursion relations is that although
initially the coefficient $a_\omega =0$, for $b>1$ $a_\omega (b)
>0$. This expresses a simple physical effect that the interaction
between two distant charges in three spacetime dimensions in a
medium with a finite polarizability is always Coulombic ($\sim
1/\sqrt{x^2 + \tau^2}$) in three spacetime dimensions
\cite{igor1}. The only apparent way the generation of this term
can be avoided would be if the propagator $G(\vec{k},\omega)$
would be {\it completely} independent of frequency, in which case
$G_> (x,\tau) \sim \ln(b) e^{-|x|\Lambda} \delta(\tau)$, rendering
the integral in $a_\omega (b)$ to be exactly zero. This unphysical
situation would correspond to an essentially two dimensional
sine-Gordon theory, which would then have the standard
Kosterlitz-Thouless transition.

  In the present case, on the other hand, we find
\begin{equation}
\frac{d\hat{a}_k}{d\ln(b)}= \frac{\hat{y}^2}{2\pi \hat{a}_k}+
O(\hat{y}^3), \label{e17}
\end{equation}
\begin{equation}
\frac{d\hat{a}_\omega}{d\ln(b)} = 2 \hat{a}_\omega +
\frac{\hat{y}^2} {4\hat{a}_k} + O(\hat{y}^3), \label{e18}
\end{equation}
where we have introduced the dimensionless combinations $\hat{y}=
y/\Lambda^2 \Omega$, $\hat{a}_k = a_k /\Omega$, and
$\hat{a}_\omega = a_\omega \Omega/\Lambda^2$, and introduced a
smooth cutoff as described in Appendix~\ref{app}.

As announced earlier, a finite coefficient $a_\omega$ becomes
generated to the second order in fugacity, and then becomes a
relevant coupling. With this term included the fugacity flows
according to
\begin{equation}
\frac{d\hat{y}}{d\ln(b)} = (2-\frac{1}{8\pi\sqrt{\hat{a}_k
\hat{a}_\omega- (1/2\ell)^2 }}) \hat{y} + O(\hat{y}^3), \label{e19}
\end{equation}
and always becomes relevant at long length scales. In Eq. (\ref{e19})
we assumed $\hat{a}_k \hat{a}_\omega  \gg (1/2\ell)^2 $; other limits
and more complete expressions are presented
in Appendix B. We interpret this as an
instability of the deconfined phase in the original theory, since
the instantons are always in the plasma phase, and the interaction
between two well-separated instantons is exponentially screened.

Previous renormalization group analyses of instanton effects in
U(1) spin liquids by Nagaosa \cite{nagaosa3}, Ioffe and Larkin
\cite{ioffelarkin}, and Wen \cite{wennew} accounted {\em only} for
the flow of the instanton fugacity in Eq. (\ref{e19}). Taken on
its own, this equation would indeed suggest that there is a regime
of parameters ($8 \pi \sqrt{\hat{a}_k \hat{a}_\omega-(1/2\ell)^2 }
< 1/2$ for Eq. (\ref{e19})) where the fugacity $y$ flows to zero,
and the instanton effects are negligible: this was the conclusion
of these earlier works \cite{nagaosa3,ioffelarkin,wennew}.
However, this conclusion does not hold after accounting for the
flows in Eqs. (\ref{e17},\ref{e18}).

\section{Conclusions}
\label{sec:conc}

The primary conclusion of this paper is that a two (spatial)
dimensional system of fermionic spinons residing on an incipient
Fermi surface, and interacting with a compact U(1) gauge force, is
generically in a confining phase at zero temperature.
Consequently, there are no free spinon-like excitations. When
combined with earlier results on the confinement of spinons with a
Dirac fermion spectrum \cite{quote,marston,murthy,sp,igor1}, our
results rule out the existence of a spin liquid phase in two
spatial dimensions with deconfined spinons interacting with a U(1)
gauge force. In particular, it appears that none of the algebraic, U(1),
and SU(2) symmetric spin liquids in Wen's classification \cite{wennew} are
stable phases of matter.

Our analysis also points to the existence of possible
(multi)-critical points at which the confinement picture does not
directly hold. These correspond to points where accidental
cancellations cause the renormalized values of $\hat{a}_k$ and/or
$\hat{a}_\omega$ to vanish. These are likely quantum critical
points between different confining phases, and not conventional
Coulomb states. Similar points appeared in analyses of quantum
dimer models \cite{rk,sondhi}.

Finally, we note that the arguments of this paper do not extend to
{\em three} spatial dimensions. Here, a compact U(1) gauge field
can be in a Coulomb phase, and interesting U(1) spin liquids are
possible \cite{d3}.

\begin{acknowledgments}
We are grateful F.~Nogueira and A.~Sudbo for a number of very
constructive and useful discussions. We also thank E. Fradkin, N. Nagaosa and
X.-G. Wen for helpful exchanges. This research was supported by
NSERC of Canada and the Research Corporation (I.H. and B.S.) and
by US NSF Grant DMR-0098226 (S.S.) and DMR-0071611 (G. M.).
\end{acknowledgments}

\appendix

\section{Screening of instanton interactions}
\label{cg}

The analysis in the body of the paper, and in the
Ref.~\onlinecite{igor1} has been carried out using the dual
sine-Gordon representation. While this is technically a useful way
to proceed, it does have the disadvantage of obscuring some of the
underlying physics. To remedy this, we present our arguments here
using the direct partition function for the instantons. For
simplicity, we will restrict our discussion here to the case of
Dirac fermion spinons, where the `relativistic' nature of the
theory does simplify the formalism by making the physics isotropic
in spacetime. However, similar arguments can be advanced also for
the non-relativistic case. The computations below expand on the
remarks in Refs.~\onlinecite{quote,marston,murthy,sp}, and were
implicitly carried out in Ref.~\onlinecite{murthy}. They also
connect with some of the arguments presented in
Ref.~\onlinecite{igor1}, as argued below.

With spinons with a Dirac fermion spectrum, two instantons have an
action which depends logarithmically on the separation between
them \cite{ioffelarkin,rantner,wennew}. Let us consider a model of
integer `charges' $m_i$ (instantons) at spacetime positions
$\vec{R}_i$, with the interaction term in the action $\sum_{i<j}
m_i m_j V(\vec{R}_i - \vec{R}_j)$, where
\begin{equation}
V(\vec{R} ) \equiv- K \ln (|\vec{R}|/a),
\end{equation}
with $K$ a coupling constant. We consider such an interaction in
$D$ spacetime dimensions. We are interested here is $D=3$, but the
general $D$ analysis will also allow us to connect with classical
results on the Kosterlitz-Thouless transition.

Following the original analysis of Kosterlitz \cite{kosterlitz},
we compute the effective interaction $V_{\rm eff} (\vec{R})$
between two well separated test charges: a $+1$ charge at
$\vec{R}_1$ and a $-1$ charge at $\vec{R}_2$. To lowest order in
the instanton fugacity, $y$, this interaction is renormalized by
instanton dipoles. In the spirit of the renormalization group, we
consider only the renormalization by dipoles consisting of
opposite charges with spacetime separation between $a$ and $a+da$.
So the dipole consists of a $+1$ charge at $\vec{R} + \vec{a}/2$
and a $-1$ charge at $\vec{R}-\vec{a}/2$. As in
Ref.~\onlinecite{kosterlitz}, to leading order in $y$ the
expression for $V_{\rm eff}$ is
\begin{eqnarray}
&& \exp \left( V_{\rm eff} (\vec{R}_1 - \vec{R}_2) \right)  =
\frac{\exp \left( V (\vec{R}_1 - \vec{R}_2) \right)}{Z} \Biggl[ 1
+ \nonumber \\
&&~~~~~y^2 \int_a^{a+da} d^D \vec{a} \int d^D \vec{R} \exp \left(
- V( \vec{R}_1 - \vec{R} - \vec{a}/2) + V( \vec{R}_1
- \vec{R} + \vec{a}/2)\right. \nonumber \\
&&~~~~~~~~~~~~~~~~~~~~~~~~~~~~~~\left.+ V( \vec{R}_2 - \vec{R} -
\vec{a}/2)- V( \vec{R}_2 - \vec{R} + \vec{a}/2)\right)+
\mathcal{O}(y^4) \Biggr]
\end{eqnarray}
where $y$ is the charge fugacity, and the normalization
\begin{equation}
Z = 1 + y^2 \int_a^{a+da} d^D \vec{a} \int d^D \vec{R}+
\mathcal{O}(y^4).
\end{equation}
The integral in $Z$ is proportional to the volume of the system:
however, this is standard in the low density expansion of any
system, and the volume dependence cancels in all physical
quantities order-by-order in $y$; the logarithm of the partition
function is extensive, and this dependence on the system volume is
merely a signal of that. In the above expression for $V_{\rm eff}$
it is assumed that any two charges are never less than a distance
$a$ apart, and this cuts off any incipient ultraviolet
divergences. Now expand the argument of the exponential in powers
of $a$. This yields
\begin{eqnarray}
&& \exp \left( V_{\rm eff} (\vec{R}_1 - \vec{R}_2) \right)  =
\frac{\exp \left( V (\vec{R}_1 - \vec{R}_2) \right)}{Z} \Biggl[1 +
\nonumber \\ &&~ y^2 \int_a^{a+da} d^D \vec{a} \int d^D \vec{R}
\exp \Biggl( - K \vec{a} \cdot \left( \frac{\vec{R}_1 -
\vec{R}}{(\vec{R}_1 - \vec{R})^2} - \frac{\vec{R}_2 -
\vec{R}}{(\vec{R}_2 - \vec{R})^2} \right) + \mathcal{O} (\vec{a}^3
) \Biggr)+ \mathcal{O}(y^4) \Biggr]
\end{eqnarray}
The integral over $\vec{a}$ can be performed, and to leading order
in $y$ and $a$ we obtain
\begin{eqnarray}
&&V_{\rm eff} (\vec{R}_1 - \vec{R}_2 ) \nonumber \\
&&~~= - K \ln (|\vec{R}_1 - \vec{R}_2|/a) + y^2 \frac{S_D K^2
a^{D+1} da}{2D} \int d^D \vec{R} \left( \frac{\vec{R}_1 -
\vec{R}}{ ( \vec{R}_1 - \vec{R} )^2} - \frac{\vec{R}_2 - \vec{R}}{
( \vec{R}_2 - \vec{R} )^2} \right)^2
\label{f1} \\
&&~~= - K \ln (|\vec{R}_1 - \vec{R}_2|/a) + y^2 \frac{S_D K^2
a^{D+1} da}{2D} \int d^D \vec{R} \frac{(\vec{R}_1 - \vec{R}_2
)^2}{(\vec{R}_1 - \vec{R})^2 (\vec{R}_2 - \vec{R})^2} \label{f2}
\end{eqnarray}
where $S_D$ is the surface area of a sphere in $D$ dimensions.
Note that the volume dependence has cancelled as expected. The
above is, of course, an adaptation of the computation performed by
Kosterlitz \cite{kosterlitz}.

Now, notice that there is a crucial distinction in the integral in
(\ref{f2}) between $D=2$ and $D=3$. In $D=2$, the integral {\em
reproduces} the functional form of the bare logarithmic
interaction, and so the renormalization of the potential can
indeed be accounted for by a dielectric constant:
\begin{equation}
V_{\rm eff} (\vec{R}_1 - \vec{R}_2 ) = - K \ln (|\vec{R}_1 -
\vec{R}_2|/a) + y^2 \frac{S_2^2 K^2 a^{3} da}{2} \left[\ln
(|\vec{R}_1 - \vec{R}_2|/a)+\mbox{constant}\right]~~,~~D=2,
\end{equation}
where the ultraviolet cutoff radius around each charge is used to
make the integral finite. This leads to the usual Kosterlitz
recursion relations \cite{kosterlitz} in $D=2$.

Turning, finally, to $D=3$, the integral over $\vec{R}$ in
(\ref{f2}) now yields
\begin{equation}
V_{\rm eff} (\vec{R}_1 - \vec{R}_2 ) = - K \ln (|\vec{R}_1 -
\vec{R}_2|/a) + y^2 \frac{S_3^2 \pi^2 K^2 a^{4} da}{24} |\vec{R}_1
- \vec{R}_2|~~,~~D=3. \label{A8}
\end{equation}
There was now no need to apply any ultraviolet cutoff. Note that
the bare logarithmic interaction is {\em not reproduced} under
renormalization--this is the central point. Instead, we have
generated a {\em linear interaction}, which has a much stronger
dependence on the separation between the charges. It is also
crucial to note the sign of this linear interaction: it is such as
to weaken the interaction between oppositely charged particles,
and at large enough $|\vec{R}_1 - \vec{R}_2|$ the effective
interaction between these opposite charges actually becomes
repulsive. Of course, we cannot trust the renormalization group
beyond these scales. Nevertheless, the lesson is quite clear: the
effect of the dipole is to substantially weaken the interaction
between the charges, suggesting a ubiquitous instability to a
plasma phase.

Continuing this line of argument, we can also look at the
screening of the generated linear interaction. Using exactly the
same derivation as above, the integral in the second term on the
right hand side of (\ref{f1}) will be replaced by
\begin{equation}
\int d^3 \vec{R} \left( \frac{\vec{R}_1 - \vec{R}}{ |  \vec{R}_1 -
\vec{R} |} - \frac{\vec{R}_2 - \vec{R}}{ | \vec{R}_2 - \vec{R} |}
\right)^2
\end{equation}
This integral is now divergent at large $\vec{R}$, which
reconfirms our conclusion that the screening effect of the largest
scales $\vec{R}$ is extremely disruptive.

  The Eq. (\ref{A8}) may also be understood as the expansion of the
screened interaction found in Ref.~\onlinecite{igor1} in small fugacity.
Considering the electrostatics of logarithmically interacting charges
bound in dipole pairs, the effective interaction due to an external
charge, in the Fourier space, becomes,
\begin{equation}
V_{\rm eff}(q) = \frac{K}{|q|^3 + \chi q^2},
\end{equation}
in $D=3$, where $\chi \sim y^2$ is the finite polarizability of the
medium \cite{igor1}. At a large distance $R$ this translates into
$V_{eff}(R) \sim 1/(4\pi \chi R)$. Expanding in small $\chi$, on the
other hand, yields
\begin{equation}
V_{\rm eff}(q) = \frac{K}{|q|^3} - \frac{K\chi}{q^4} + O(\chi^2).
\end{equation}
The first term in the expansion then represents the bare
logarithmic interaction in Eq. (\ref{A8}), while the second one
may be identified with the next, linear correction.

\section{Computation for renormalization analysis}
\label{app}

 Here we discuss the smooth cutoff procedure we devised to compute
 $G_>(x,\tau)$ in Eq. (\ref{e16}), and to arrive at the flow equations
 (\ref{e17}-\ref{e19}). With $a_\omega =0$ in the unrenormalized theory,
\begin{equation}
G_> (x,\tau) = \frac{1}{2\pi e^2}\int_{\Lambda/b}^{\Lambda}
\frac{dk}{k} J_0 (kx) \int_{-\Omega}^{\Omega} \frac{d\omega}{2\pi}
\frac{e^{i\omega\tau}} {1+ |\omega|/(e^2 \ell) }.
\label{B1}
\end{equation}
First, we write
\begin{equation}
\int_{\Lambda/b}^{\Lambda} \frac{dk}{k} J_0 (kx)=
\int_{\Lambda/b}^{\infty} \frac{dk}{k} J_0 (kx)
-\int_{\Lambda}^{\infty} \frac{dk}{k} J_0 (kx)
\end{equation}
and then introduce a smooth cutoff \cite{igor1} as
\begin{equation}
\int_{\Lambda}^{\infty} \frac{dk}{k} J_0 (kx)
\rightarrow \int_{0}^{\infty} \frac{dk}{k}J_0 (kx)\frac{k^2}{k^2 + \Lambda^2}.
\end{equation}
  This leads to
\begin{equation}
\frac{1}{2\pi e^2}\int_{\Lambda/b}^{\Lambda} \frac{dk}{k} J_0 (kx)\rightarrow
-\frac{x\Lambda \ln(b)}{2\pi e^2} \frac{d K_0 (z)}{dz}|_{z=x\Lambda},
\end{equation}
with $K_0 (z)$ as the Bessel function. Similarly, for $\Omega/e^2
\ell \ll 1$ we may neglect the frequency dependence and
approximate the second integral in Eq. (\ref{B1}) as
\begin{equation}
\int_{-\Omega}^{\Omega} \frac{d\omega}{2\pi} e^{i\omega\tau} \rightarrow
\int_{-\infty}^{\infty} \frac{d\omega}{2\pi} e^{i\omega\tau}
e^{-\pi(\omega/(2\Omega))^2},
\end{equation}
where we have introduced the Gaussian  instead of
the sharp cutoff in the frequency space. This gives
\begin{equation}
\int_{-\Omega}^{\Omega} \frac{d\omega}{2\pi}
\frac{e^{i\omega\tau}} {1+|\omega|/(e^2 \ell ) } \rightarrow
\frac{\Omega}{\pi} e^{-(\tau \Omega)^2 /\pi } .
\end{equation}
Inserting the last two expressions into Eqs. (\ref{e13}) and
(\ref{e14}) yields the $\sim y^2$ terms in the recursion relations
Eqs. (\ref{e17}) and (\ref{e18}).

  Assuming $a_\omega >0$, the frequency cutoff $\Omega$ may be taken
to infinity in the computation  of $G_> (0,0)$ appearing in the
Eq. (\ref{e15}). This leads to
\begin{equation}
\frac{d\hat{y}}{d\ln(b)} = \left[ 2-\frac{\ell}{4\pi D^{1/2} }
\left( 1-\frac{2}{\pi} \arctan \left( \frac{1}{D^{1/2}}\right)
\right) \right] \hat{y} + O(\hat{y}^3),
\end{equation}
for $D>0$, where
\begin{equation}
D= 4 \hat{a}_k \hat{a}_\omega \ell ^2- 1.
\end{equation}.

  For $D<0$, one finds
\begin{equation}
\frac{d\hat{y}}{d\ln(b)} = \left[ 2-\frac{\ell}{ 4 \pi^2
|D|^{1/2}} \ln\left(\frac{1+ |D|^{1/2} }{1-|D|^{1/2}} \right)
\right] \hat{y} + O(\hat{y}^3).
\end{equation}

For completeness, let us also mention the case $a_\omega (b) \equiv 0$,
which would
arise if one would neglect  $\sim y^2$ terms in the recursion relations.
One then  needs to keep a finite cutoff $\Omega$, to find
\begin{equation}
\frac{d\hat{y}}{d\ln(b)} = \left[ 2-\frac{\ell}{4 \pi^2} \ln
\left(1+ \frac{1}{\ell \hat{a}_k} \right) \right] \hat{y} +
O(\hat{y}^3). \label{b10}
\end{equation}
Because to the lowest order in $y$ neither $\ell$ nor $\hat{a}_k$
scale, the last equation would suggest that fugacity is irrelevant
for small enough $\hat{a}_k$, for example. This conclusion,
however, inevitably breaks down to the next order in $y$, as
discussed in the body of the paper.

  Note also the similarity between the Eq. (\ref{b10}) and Eq. (11) in
 Ref.~\onlinecite{nagaosa3}, upon identification of $\ell$ with $\gamma$, and
 $a_k$ with $g^2$. The main difference is that in Nagaosa's
 renormalization scheme the coupling $g^2$ (analogous to our
 coefficient $a_\omega$)
 appears {\it irrelevant } by naive power counting,
 which then suggest that fugacity   should
 become irrelevant as well. This conclusion, however, changes to the
 next order in fugacity, in which the fugacity always becomes relevant
even within Nagaosa's renormalization
 scheme, and much like in our calculation.


\begin{thebibliography}{}
\bibitem{nagaosanote} In his analysis of the spinon Fermi surface, Nagaosa
\protect\cite{nagaosa3} presents a separate argument for a
deconfined phase in the first part of his paper. By a
decomposition of the gauge field into its longitudinal and
transverse parts, he reduces the theory to an effective XY model
with dissipation. However, this approximation leads to a model
with only line vortices, and so also effectively neglects
instanton events at points in spacetime.
\bibitem{comment}  In Ref.~\protect\onlinecite{nagaosa3},
instead of the $\sim k^2$ term in our Eq. (\protect\ref{e10}),
Nagaosa finds a regular $\sim \omega^2$ term. He then chooses a
renormalization procedure complementary to ours, in which modes
with $\Omega/b < |\omega| < \Omega$ and with all $k< \Lambda$ are
integrated out. Although the origin of this difference in the
sine-Gordon actions is not clear to us, it is easy to see that
even within Nagaosa's scheme a regular $\sim k^2$ term gets
generated to the second order in fugacity, and this ultimately
always makes the fugacity relevant. See also the discussion at the
end of Appendix~\protect\ref{app}.
\end{thebibliography}
\end{document}